%% file: Main.tex
\documentclass{article}
\usepackage{spconf,amsmath,graphicx}
\usepackage{amssymb}
\usepackage{subcaption}
\usepackage{algorithm}
\usepackage{algpseudocode}
\usepackage[table,xcdraw]{xcolor}

\title{Distributed Injection-Locking in Analog Ising Machines to Solve Combinatorial Optimizations}
\name{M. Ali Vosoughi\thanks{Accepted to ISCAS 2020, this is a preprint version in \cite{vosoughi2020distributed}.}}
\address{Department of Electrical and Computer Engineering, University of Rochester, NY, USA}
\begin{document}
\ninept
\maketitle
\begin{abstract}
The oscillator-based Ising machine (OIM) is a network of coupled CMOS oscillators that solves combinatorial optimization problems. 
In this paper, the distribution of the injection-locking oscillations throughout the circuit is proposed to accelerate the phase-locking of the OIM. 
The implications of the proposed technique theoretically investigated and verified by extensive simulations in EDA tools with a $130~nm$ PTM model.
By distributing the injective signal of the super-harmonic oscillator, the speed is increased by $219.8\%$ with negligible increase in the power dissipation and phase-locking error of the device due to the distributed technique.
\end{abstract}

\input{sections}


\bibliographystyle{IEEEbib}
\newpage
\bibliography{strings,refs}

\end{document}

%% file: sections.tex
\section{Introduction}
With the advent of world-wide communications and the availability of massive data, increasing application of neural networks and artificial intelligence and end of Moore's law, computing becomes a significant obstacle in recent decades~\cite{lecun2015deep,bojnordi2016,longfei_fog,ali_combined}. 
Particularly computational power is crucial in real-time and high-dimensional machine learning settings, such as detecting anomalous processes~\cite{mozaffari2019online1,mozaffari2019online2}.
The Ising model has attracted considerable prominence due to its network-based structure and the ability to map many NP-complete problems to Ising model, and various accelerators and machines have been proposed~\cite{de2016simple,lucas2014ising}.
Remarkable endeavors to direct the computational demands of optimization incorporates quantum, optical, and emerging non-Von Neumann computing paradigms. 

Quantum and optical techniques are deemed competent to address projected markets for ever-increasing computational demands. 
The D-wave technique that proposes optimization by qubits~\cite{johnson2011quantum} and the optical technique based on the coherent Ising machine (CIM) that performs the light-based computations are some of the state-of-the-art proposed techniques~\cite{inagaki2016coherent}.
Notwithstanding the high computational capabilities of these techniques, their practical utilization is quite expensive and ill-suited for commercialization~\cite{johnson2011quantum, inagaki2016coherent}.
Alternatively, inexpensive CMOS-based techniques, such as the oscillator-based Ising machine (OIM) and the CMOS annealer, have been proposed~\cite{wang2019oim,annealer2019isscc,chou2019analog} .  
Analog architectures have manifested thousands of times increase in the performance of computation as compared to digital competitors, thanks to their dynamic accuracy~\cite{darpa2019ai_campain,analog_meet_digital}. 
Computation with analog architectures is one of the key plans that DARPA has featured in the artificial intelligence next campaign~\cite{darpa2019ai_campain}.

%

The OIM technique is an analog CMOS optimization accelerator that is based on a population of coupled CMOS oscillators~\cite{wang2019oim}.
In an OIM, by mapping the combinatorial optimization to the circuit using a mapping interface array circuit, attractive and repulsive connections are formed between the oscillators, which leads to a new order, called oscillator glass~\cite{glass1991daido}. 
Then, the phases that are near zero or $\pi$ are measured and processed to resolve the optimization problem, which conforms to the Ising model.

The phases of the oscillatory network of OIM can spontaneously drift, or lock with a lag or lag as compared to the overall phase of the oscillator population. 
Accordingly, an external injection-locking oscillator is used to align the oscillator phases of the OIM to near zero and $\pi$.
Injection-locking is a technique utilized for phase-locking of a CMOS oscillator~\cite{adler1947}. 
In OIM~\cite{wang2019oim,chou2019analog}, injection-locking has been used to lock the phases of oscillators to $\{0,\pi\}$.
Authors of \cite{wang2019oim} propose the use of injection-locking on various populations of oscillators, and in \cite{chou2019analog} a mapping interface array circuit is proposed to conform a max-cut problem to an OIM ranging from hundreds to thousands of coupled oscillators, without detailing the method to address the injection-locking. 
Regrettably, there is no attempt in the literature to investigate the implications of various implementations for injection-locking techniques on the operation of an OIM device. 
We investigate the implications of centralized and distributed injection-locking techniques on the operation of the OIM circuit and propose the distributed injection-locking signals through the OIM to enhance the speed of phase-locking of the oscillators.
We show that by distributed injection-locking technique, the optimization speed is increased by $219.8\%$ with negligible power and phase-locking error overhead as compared to the centralized counterpart.

The rest of the paper is as follows. 
In section \ref{sec:background}, the background for the Ising model and its mapping to the max-cut problem is explained. 
In section \ref{sec:proposed} the proposed technique is analyzed and modeled, followed by simulations in section \ref{sec:simulations}. 
The paper is concluded in section \ref{sec:conclusion}. 
The present paper is in line with the author's measures to address emerging technologies and concerns in modern integrated circuits \cite{vosoughi201610,vosoughi2017noise,vosoughi2019bus,ali_combined,vosoughi2019leveraging, vosoughi2020distributed}.

\section{Background}\label{sec:background}
\begin{figure}[!h]
\centerline{\includegraphics[width=0.45\textwidth]{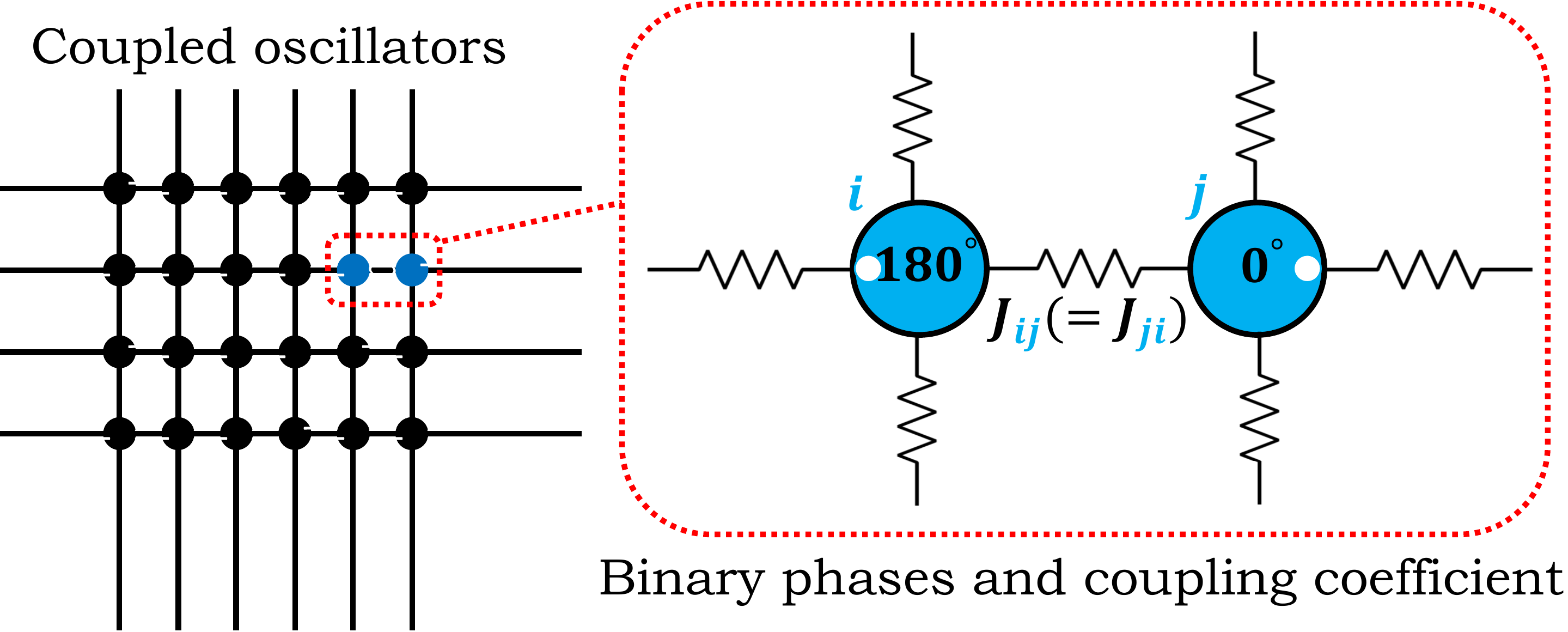}}
\caption{The CMOS oscillators are interconnected in the Ising network and take binary phases $0^{\circ}$ or $180^{\circ}$ under the influence of network-wide coupling coefficients $J_{ij}$. The binary phases are the solution to the problem of max-cut optimization}
\label{fig:IsingMachine}
\end{figure}

The Ising model is to study the phase transition from disorder to order for Hamiltonian of a system.
The Ising model is defined for a one-dimensional, two-dimensional, or multi-dimensional networks, such that at every point $i$,  there is a spin $S_i$ which takes only two values of $+1$ and $-1$.
The name "spin" originates from the model's initial application for investigating order in a magnetic material.
Hamiltonian of a system is defined as follows,
\begin{equation}
H=-\sum_{(i,j)} J_{ij}S_iS_j - \sum_i h_iS_i,
\end{equation}
where the sum is over the neighboring spins. 
In the Hamiltonian, $h_i$ represents an external field, and $J_{ij}$ represents the spin-spin interaction.
For negative $J_{ij}$,  the nearby spins take the opposite direction, and for positive $J_{ij}$, take the same direction to ground the Hamiltonian to reach its minimum state.
According to the second law of thermodynamics, every system naturally tends to minimize energy.
In the absence of an external field, $h_i$ equals zero, and the Ising model is simplified as follows,
\begin{equation}
\begin{aligned}
\min_{\bold{S}} \quad & H=-\sum_{(i,j)} J_{ij}S_iS_j\\
\textrm{s.t.} \quad &  S_i\in \{-1,+1\}  ,\\
  \quad &  \forall \{i\neq j\} \in \{1,2,\dots,N\}  ,\\
  \quad &  J_{ij}=J_{ji}.  
\end{aligned}
\end{equation}
The Ising model can be mapped to the equivalent graph model where the vertices in the graph can take two values $\{\pm 1\}$, and the weights of the edges are equivalent to the coupling coefficient $J_{ij}$ between the spins. 
Such a graph is equivalent to the graph of the max-cut optimization problem where the positive and negative spin of the corresponding vertex indicates which side of the cut a particular vertex lies on.
By replacing the vertices with CMOS oscillators and the graph edges with proper interconnects, the optimization machine based on the Ising model, as depicted in Fig. \ref{fig:IsingMachine}, is realized.
The phase of these oscillators is equivalent to the spin in the Ising model, which is equal to the solution of the max-cut problem~\cite{wang2019oim,chou2019analog}.

\section{Distributed injection-locking}\label{sec:proposed}
\begin{figure}[!h]
\centerline{\includegraphics[width=0.45\textwidth]{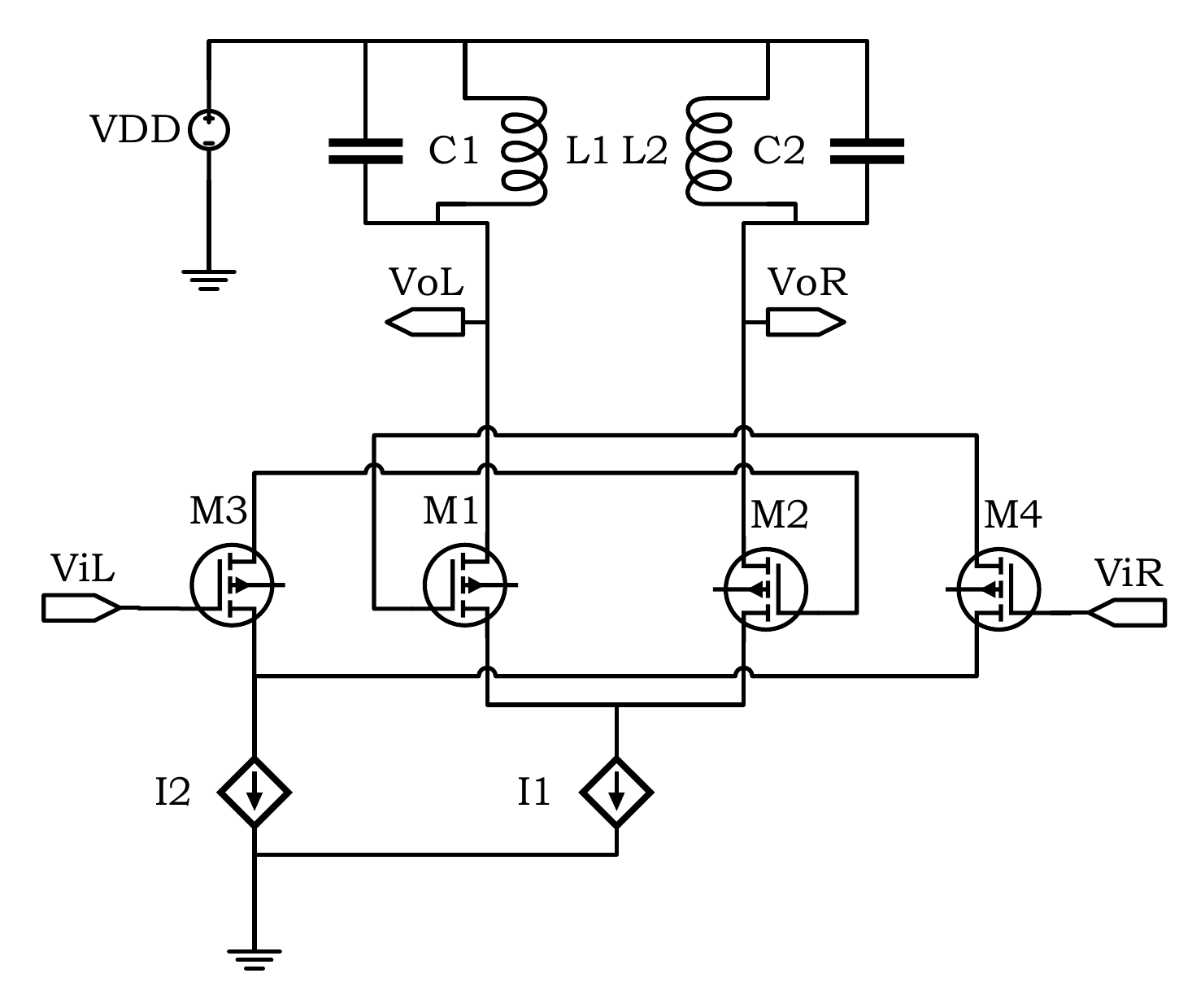}}
\caption{The oscillator of OIM, proposed in \cite{chou2019analog}. The oscillator has differential input to receive the coupling coefficients of the max-cut problem $J_{ij}$. The current $I_1$ supplies the current of each oscillator, while current $I_2$ is the signal of external injection-locking oscillator.}
\label{fig:cmos_oscillator_MIT}
\end{figure}

Without loss of generality, we use the CMOS oscillator that is proposed by \cite{chou2019analog} to investigate the implications of centralized and distributed injection-locking techniques on the operation of the OIM, as shown in Fig. \ref{fig:cmos_oscillator_MIT}. 
It is worth mentioning that the injection-locking signal $f_{inj}$ impacts the OIM by \textit{Adler} equation~\cite{adler1947}.
The injection-locking signal $f_{inj}$ is applied by $I_{2}$ to the oscillators of the OIM (Fig. \ref{fig:cmos_oscillator_MIT}).
In the centralized technique of injection-locking, despite a large number of oscillators of OIM, \textit{i.e.,} thousands of oscillators, there is a current $I_{2,j},\forall j$ for each oscillator. 

\begin{figure}[!h]
\centerline{\includegraphics[width=0.45\textwidth]{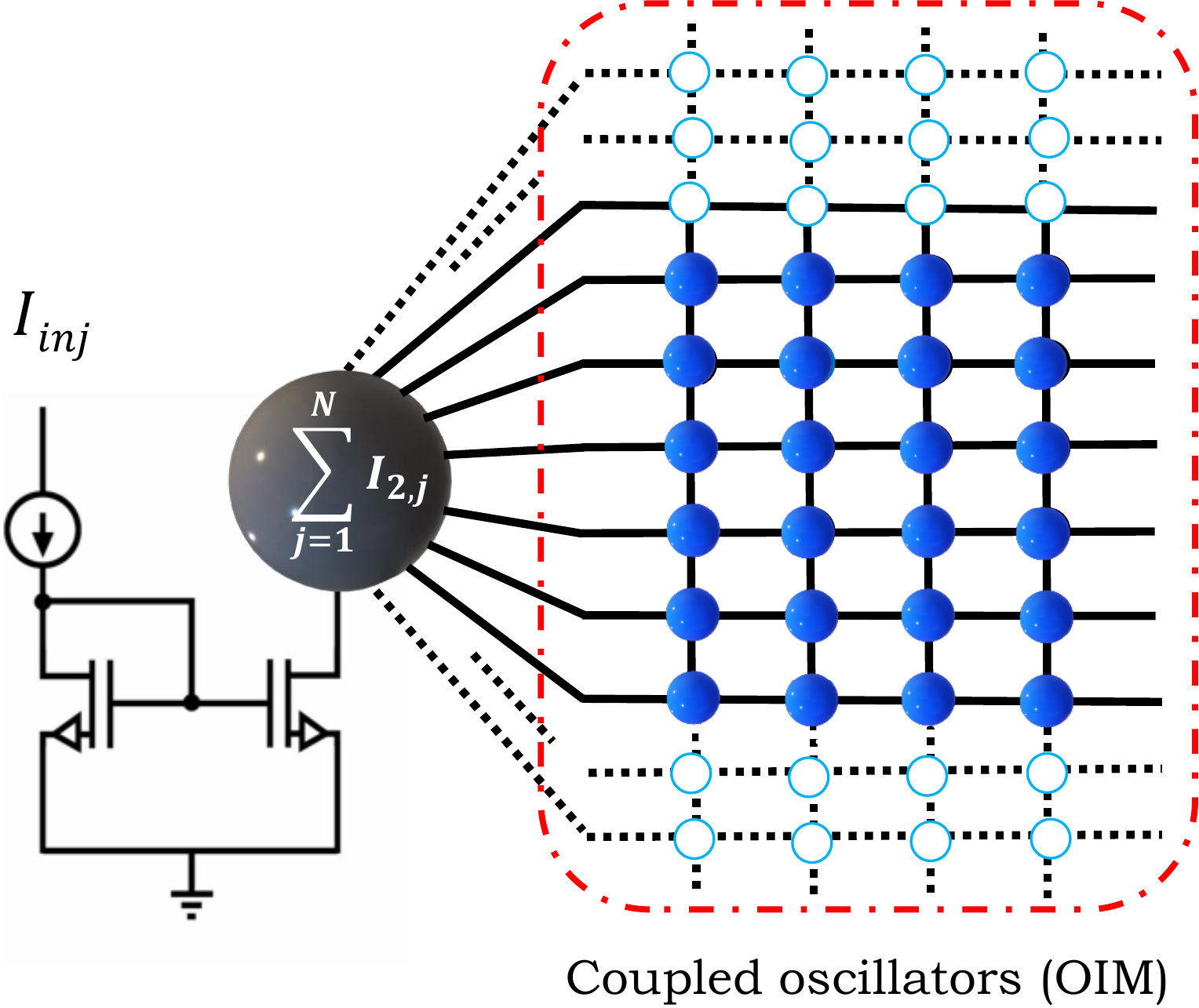}}
\caption{In centralized injection-locking, all of the oscillators of the OIM are connected to one global injection-locking signal by a shared routing. The shared routing $\sum^N_{j=1}I_{2,j}$ induces interference among the oscillators of OIM, ending in performance degradation of the OIM.}
\label{fig:centralized_SHIL_micro}
\end{figure}

\begin{figure}[!h]
\centerline{\includegraphics[width=0.45\textwidth]{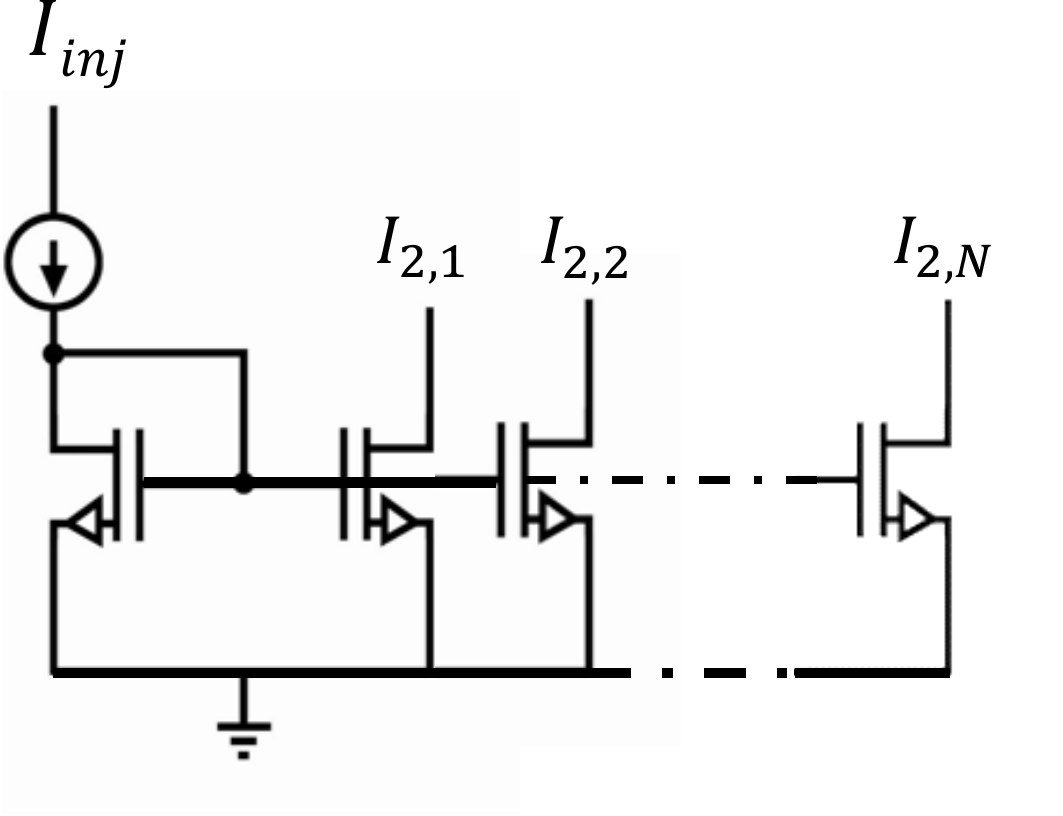}}
\caption{The proposed technique for the distribution of the injection-locking signal to oscillators of the Ising machine to enhance the performance of the OIM. Distributed current mirrors are used to reduce inter-oscillator interference. }
\label{fig:distributed_SHIL_micro}
\end{figure}

If the $I_2$ of oscillators are coupled to the injection-locking oscillator via a shared routing, as shown in Fig. \ref{fig:centralized_SHIL_micro}, the centralized injection-locking will follow two behaviors that are as follows.
The OIM is composed of many oscillators, where each one oscillates at frequency $\omega_i$. 
By coupling of the identical oscillators, the interaction between the oscillators $J_{ij}$ (as explained in Section \ref{sec:background}), moves the population from a disorder to the new order, called glass oscillator~\cite{glass1991daido}.
Consequently, the new order in the oscillator population can be written as~\cite{glass1991daido},
\begin{equation}\label{eqn:kuramoto}
\Dot{\theta_i}(t)=\omega_i-\sigma\sum_{j=1}J_{ij}sin(\theta_i(t)-\theta_j(t)),
\end{equation}
which is a generalized form of the Kuramoto model.
In \eqref{eqn:kuramoto}, $\omega_i$ is the frequency of an uncoupled oscillator, $\sigma$ is the strength of coupling, and $J_{ij}$ is the coupling coefficient, as previously explained.
By increasing $\sigma$ to above a particular threshold $\sigma_c$, the oscillators of the OIM synchronize~\cite{glass1991daido}.
As the oscillators are synchronized, the frequency $\omega_i$ of each oscillator converges to the mean-field frequency ($\omega_i\rightarrow \Omega,~\forall{i}$), letting the $\omega_i$s be eliminated, which yields the relationship,
\begin{equation}\label{eqn:kuramoto_no_w}
\Dot{\theta_i}(t)=-\sigma\sum_{j=1}J_{ij}sin(\theta_i(t)-\theta_j(t)).
\end{equation}
By coupling injection-locking signal $\theta_{inj,j},\forall j$  to each oscillator of the OIM with coupling strength $\kappa_s $, the first behavior emerges as,
\begin{equation}\label{eqn:kuramoto_with_SHIL}
\Dot{\theta_i}(t)=-\sigma\sum_{j=1}J_{ij}sin(\theta_i(t)-\theta_j(t))-\kappa_s{sin(\theta_{inj}(t))},
\end{equation}
which is desired.
However, the oscillators are coupled to injection-locking oscillator by a shared routing $I_{inj}=\sum^N_{j=1}I_{2,j}$, driving a different phase dynamics,
\begin{equation}\label{eqn:kuramoto_with_SHIL_dynamics}
\Dot{\theta_i}(t)=-\kappa_s\sum_{j=1}sin(\theta_{i}(t)-\theta_{j}(t))-\kappa_s{sin(\theta_{i}(t)-\theta_{inj}(t))}.
\end{equation}
Accordingly, the phase dynamics of the oscillator becomes,
\begin{equation} \label{eqn:kuramoto_deviated}
\begin{aligned}
&\Dot{\theta_i}(t)=-\kappa_s\sum_{j=1}sin(\theta_{i}(t)-\theta_{j}(t))
-\kappa_s{sin(\theta_{i}(t)-\theta_{inj}(t))} \\
&-\sigma\sum_{j=1}J_{ij}sin(\theta_i(t)-\theta_j(t))-\kappa_s{sin(\theta_{inj}(t))}.
\end{aligned}
\end{equation}
Expression \eqref{eqn:kuramoto_deviated} is an ill-suited phase dynamics for an OIM. 
To solve the problem of dynamics of \eqref{eqn:kuramoto_deviated}, the technique of distribution of the injection-locking currents, as shown in Fig. \ref{fig:distributed_SHIL_micro}, is proposed.
The proposed technique, as shown in Fig. \ref{fig:distributed_SHIL_micro}, incorporates distributed current mirrors to reduce inter-oscillator interference.
The Kuramoto model for the OIM with distributed injection-locking currents will be,
\begin{equation}\label{eqn:kuramoto_with_SHIL}
\Dot{\theta_i}(t)=-\sigma\sum_{j=1}J_{ij}sin(\theta_i(t)-\theta_j(t))-\kappa_s{sin(\theta_{inj}(t))},
\end{equation}
that states the phase dynamics for the Kuramoto model in an Ising machine \cite{wang2019oim}.
%

%

\section{Simulations}\label{sec:simulations}
\begin{figure}[!h]
\centerline{\includegraphics[width=0.45\textwidth]{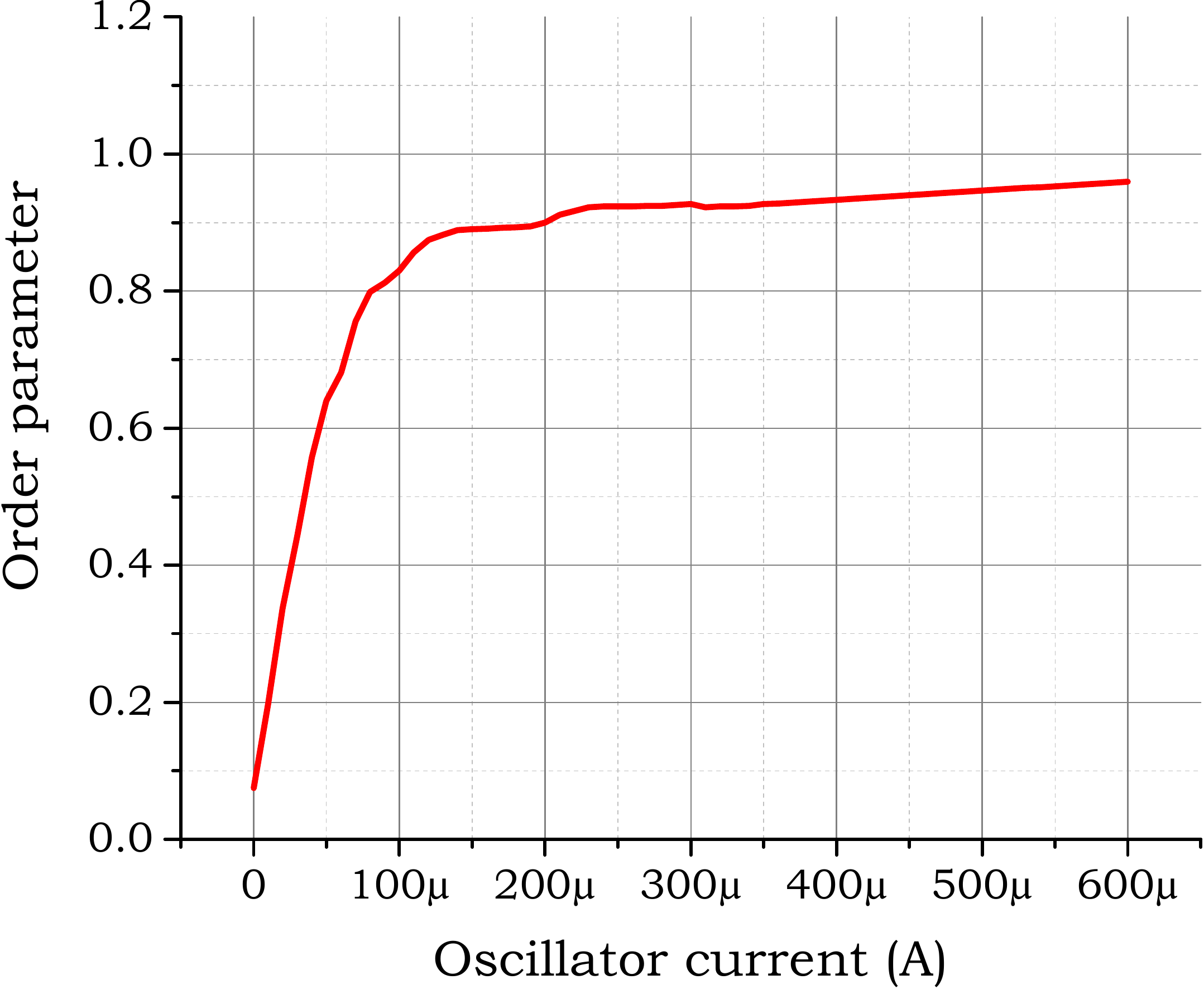}}
\caption{The variation of the order parameter $R$ of the coupled oscillators for various currents of the oscillators $I_1$ is shown.  By increasing the current of the oscillators, the order parameter $R$ of the coupled oscillators increases, and no phase-locking occurs for currents below a certain limit of $I_1$.}
\label{fig:sweep_coupling}
\end{figure}

The simulations are performed in EDA tools (Virtuoso Cadence and ADS Keysight) with $130~nm$ PTM model for bulk CMOS of the predictive technology model\cite{ptm}.
Ten coupled oscillators with the frequency of  $50.3MHz$, the inductor $100nH$, and the capacitor $100pF$ are implemented and coupled with resistive interconnects. 
An ideal current source at frequency $100.6MHz$ provides the current to the oscillators through centralized (Fig.~\ref{fig:centralized_SHIL_micro}) and distributed (Fig.~\ref{fig:distributed_SHIL_micro}) current mirrors. 
Order parameter $R$ is defined as~\cite{frasca2018synchronization},
$
R(t)=\frac{1}{N}|\sum_{i=1}^Ne^{j2\theta_i(t)}|
$
to define phase-locking order of the $N$ oscillators $\theta_i, \forall i$ to the in-phase and anti-phase synchronizations of $\{0,\pi\}$.
As shown in Fig. \ref{fig:sweep_coupling}, by increasing the current of the oscillators (increasing $I_1$ of Fig. \ref{fig:cmos_oscillator_MIT}), the order parameter $R$ of the oscillators increases. 
An interesting result of variations of $R$ is that practically no phase-locking occurs for currents below a certain limit of $I_1$.
This means that by increasing the power of the OIM, it is plausible to moderately accelerate the optimization tasks running on the OIM, though after a while the enhancement becomes limited.

\begin{figure}[!h]
\centerline{\includegraphics[width=0.45\textwidth]{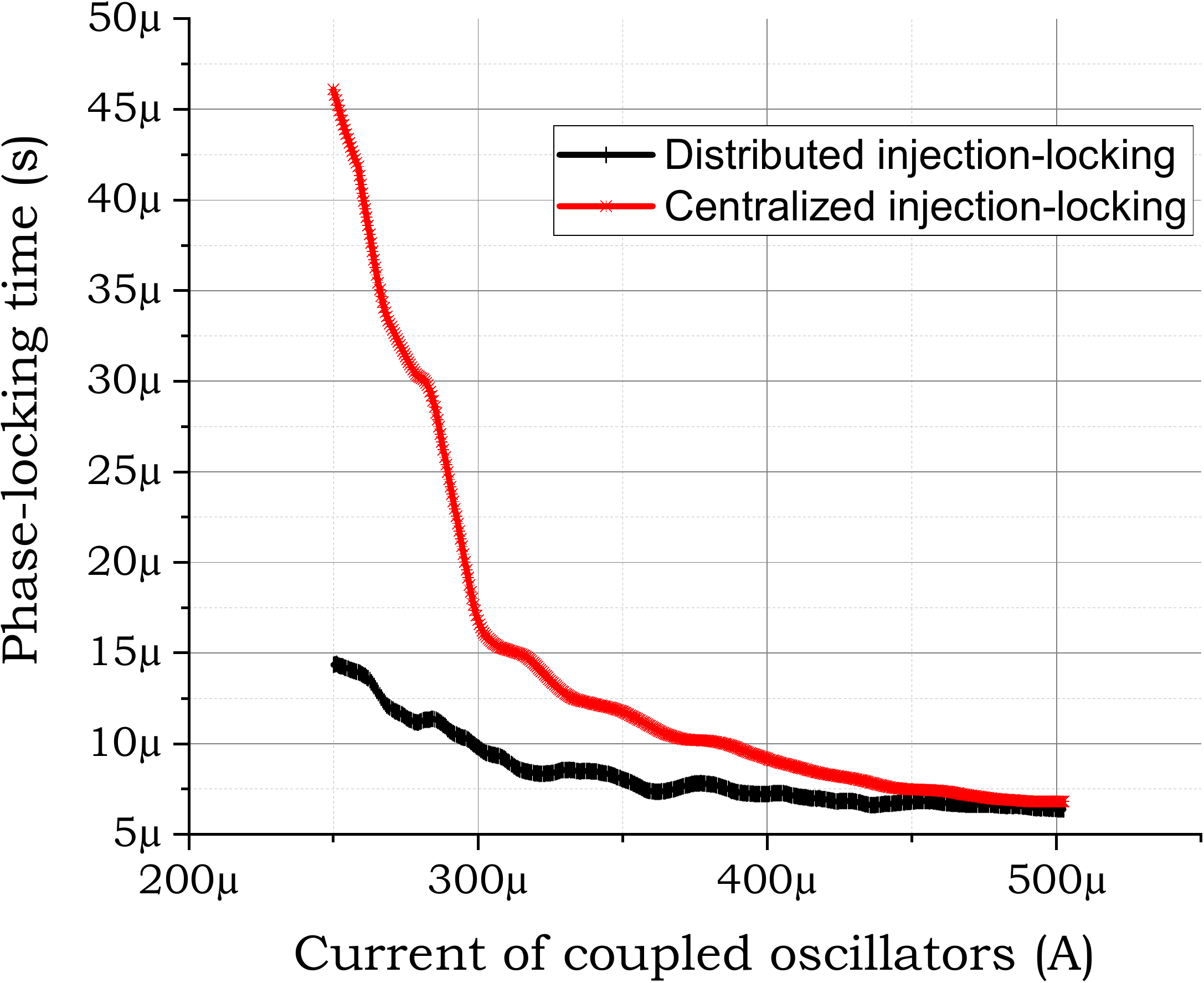}}
\caption{Effect of increasing the power of the oscillator $I_1$ to the time required for phase-locking of the coupled oscillators. Using the distributed current mirrors decreases the phase-lock time of the oscillators from $45.31~\mu s$ for centralized injection, to $14.17~\mu s$ for distributed current mirrors.}
\label{fig:sweep_current}
\end{figure}

\begin{figure}[!h]
\centerline{\includegraphics[width=0.45\textwidth]{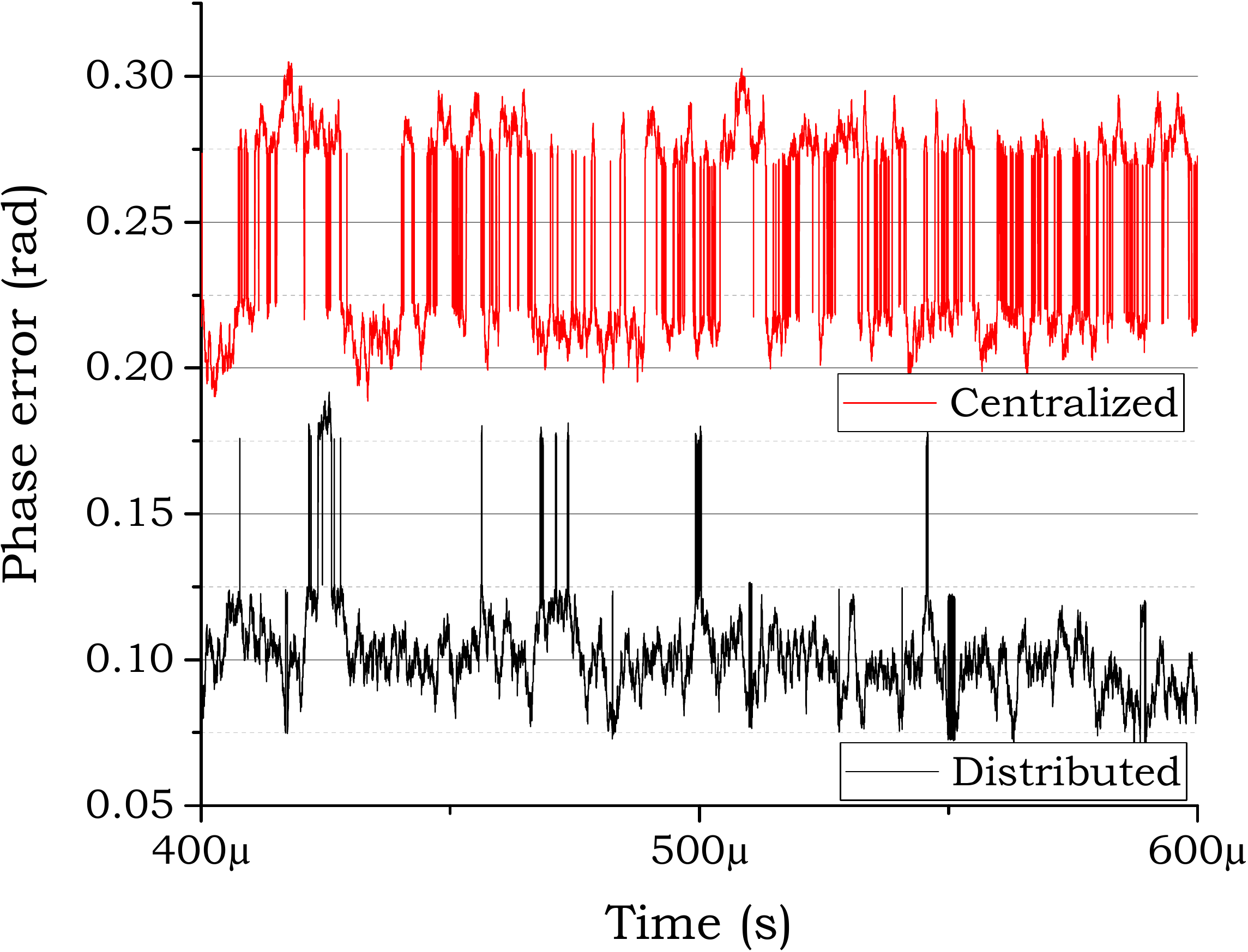}}
\caption{Total phase-locking error for distributed and centralized techniques for injection-locking simulations of Kuramoto model.}
\label{fig:phase_lock_error_total}
\end{figure}

Fig. \ref{fig:sweep_current} shows the effect of increasing the power of the injection-locking oscillator $I_2$ to the time required for phase-locking of the OIM.
As shown in Fig. \ref{fig:sweep_current}, with increasing the current $I_1$, for both distributed and centralized injection-locking, the time required for phase-locking is reduced, but this decrease is marginal after a certain amount of current.
In Fig. \ref{fig:sweep_current}, using the distributed current mirrors decreases the phase-lock time of the oscillators from $45.31~\mu s$ for centralized injection, to $14.17~\mu s$ for distributed current mirrors at $\sum^{10}_{i=1}{I_{2,i}}=1 mA$, improving the phase-locking speed of the oscillators by $219.8\%$. 
But the contrast in the benefits of distributed and centralized current injection-locking is marginal by increasing the current $I_2$  as the competitive situation between the oscillators for the injection-locking current demises.
Phase-lock error is defined as,
\begin{equation}\label{eqn:error_phase_lock}
e_{\theta}(t)=\sqrt{\frac{2}{N-1}\sum_i||\theta_i(t)-\Bar{\theta(t)}||^2},
\end{equation}
where $\Bar{\theta(t)}=\frac{1}{N}\sum_j\theta_j(t)$ is the mean-field phase of the oscillators~\cite{frasca2018synchronization}.
The phase-locking of oscillators with the distributed and centralized injection-locking technique is shown in Fig. \ref{fig:phase_lock_error_total}. 
As shown in Fig. \ref{fig:phase_lock_error_total}, the phase-locking error of Kuramoto model is reduced by $53.6\%$ around the average phase-locking error for distributed injection-locking technique as compared to the centralized counterpart, while unlike the synchronization time reduction, there is no general pattern for error reduction in the simulations.

\begin{figure}[!h]
\centerline{\includegraphics[width=0.45\textwidth]{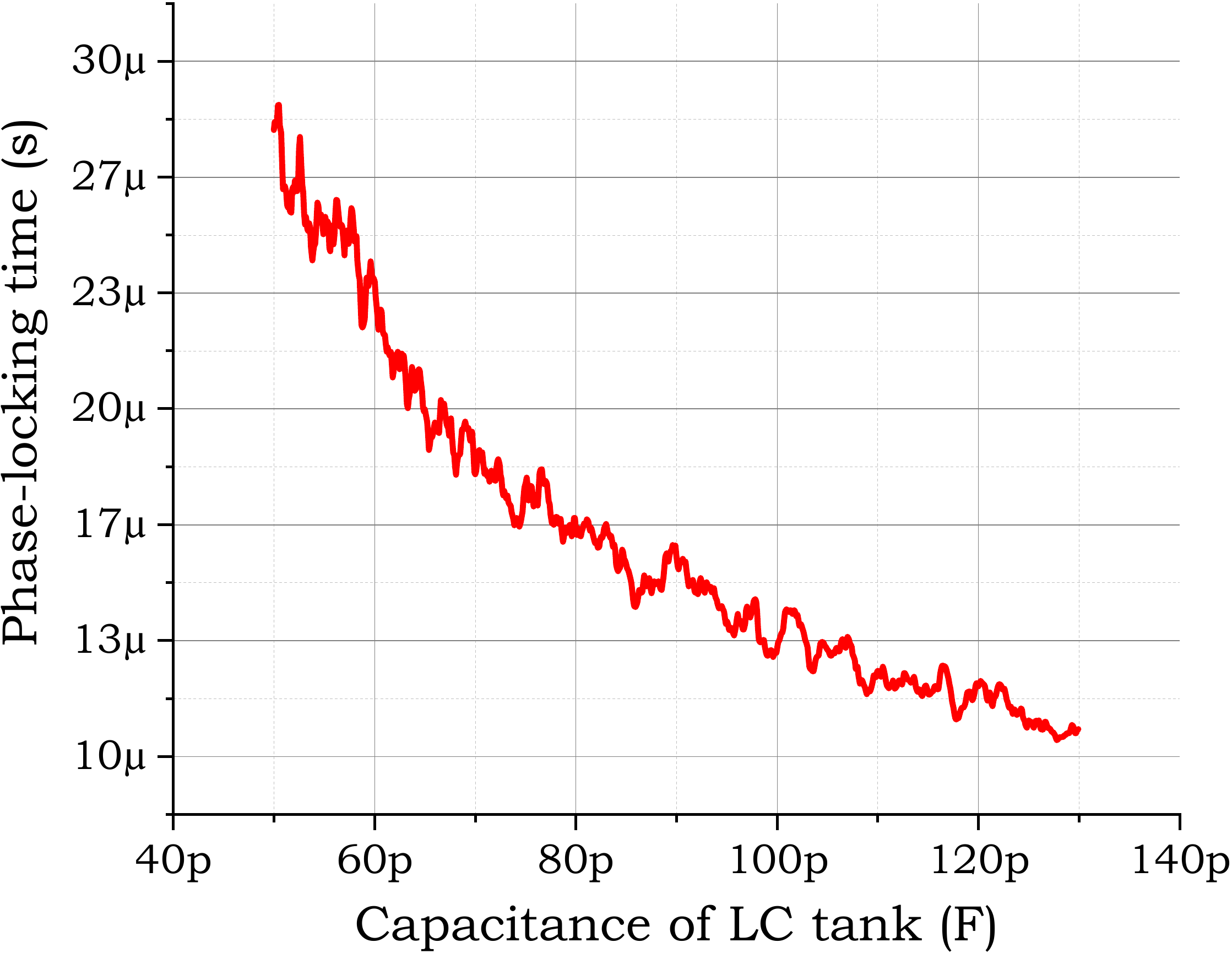}}
\caption{Scaling effect of capacitor value in the LC tank of oscillators of an OIM machine with ten coupled oscillators. Increasing the capacitor value reduces the synchronization time of the oscillators. }
\label{fig:sweep_capacitor}
\end{figure}

With an increasing capacitor size of $C$ (and similarly $L$), as shown in Fig. \ref{fig:sweep_capacitor},  the phase-locking time reduces. 
Finally, the total power of the OIM with distributed injection-locking technique is $7.188~mW$, which is $34~\mu W$ more than the power of the centralized one, implying less than $1\%$ increase in the power consumption of the device due to the distributed technique.

%


\section{Conclusion} \label{sec:conclusion}
A distributed injection-locking technique is proposed to expedite the optimization of the combinatorial problems using an OIM circuit. 
The phase dynamics of the oscillators with centralized and distributed injection-locking signals are analyzed and verified by extensive simulations.
The total power of the OIM with distributed injection-locking technique is increased by only less than $1\%$ as compared to the centralized counterpart, while the phase-locking speed of the oscillators is increased by $219.8\%$.

%% file: Main.bbl
\begin{thebibliography}{10}

\bibitem{vosoughi2020distributed}
M.A. Vosoughi,
\newblock ``Distributed injection-locking in analog ising machines to solve
  combinatorial optimizations,''
\newblock in {\em IEEE International Symposium on Circuits and Systems}. IEEE,
  2020, pp. 1--5.

\bibitem{lecun2015deep}
Y.~LeCun, Y.~Bengio, and G.~Hinton,
\newblock ``Deep learning,''
\newblock {\em Nature}, vol. 521, no. 7553, pp. 436--444, May 2015.

\bibitem{bojnordi2016}
M.~N. {Bojnordi} and E.~{Ipek},
\newblock ``Memristive boltzmann machine: A hardware accelerator for
  combinatorial optimization and deep learning,''
\newblock in {\em 2016 IEEE International Symposium on High Performance
  Computer Architecture}, March 2016, pp. 1--13.

\bibitem{longfei_fog}
L.~{Wang} and S.~{Kose},
\newblock ``When hardware security moves to the edge and fog,''
\newblock in {\em IEEE 23rd International Conference on Digital Signal
  Processing}, Novenber 2018, pp. 1--5.

\bibitem{ali_combined}
M.A. {Vosoughi} and S.~{Kose},
\newblock ``Combined distinguishers to enhance the accuracy and success of side
  channel analysis,''
\newblock in {\em IEEE International Symposium on Circuits and Systems}, May
  2019, pp. 1--5.

\bibitem{mozaffari2019online1}
M.~Mozaffari and Y.~Yilmaz,
\newblock ``Online multivariate anomaly detection and localization for
  high-dimensional settings,''
\newblock {\em arXiv preprint arXiv:1905.07107}, 2019.

\bibitem{mozaffari2019online2}
M.~Mozaffari and Y.~Yilmaz,
\newblock ``Online anomaly detection in multivariate settings,''
\newblock in {\em 2019 IEEE 29th International Workshop on Machine Learning for
  Signal Processing (MLSP)}. IEEE, 2019, pp. 1--6.

\bibitem{de2016simple}
G.~De las Cuevas and T.~S. Cubitt,
\newblock ``Simple universal models capture all classical spin physics,''
\newblock {\em Science}, vol. 351, no. 6278, pp. 1180--1183, March 2016.

\bibitem{lucas2014ising}
A.~Lucas,
\newblock ``Ising formulations of many np problems,''
\newblock {\em Frontiers in Physics}, vol. 2, pp. 5, February 2014.

\bibitem{johnson2011quantum}
M.~W.~Johnson \textit{et al.},
\newblock ``Quantum annealing with manufactured spins,''
\newblock {\em Nature}, vol. 473, no. 7346, pp. 194, May 2011.

\bibitem{inagaki2016coherent}
Takahiro~\textit{et al.} T.~Inagaki,
\newblock ``A coherent ising machine for 2000-node optimization problems,''
\newblock {\em Science}, vol. 354, no. 6312, pp. 603--606, 2016.

\bibitem{wang2019oim}
T.~Wang and J.~Roychowdhury,
\newblock ``Oim: Oscillator-based ising machines for solving combinatorial
  optimisation problems,''
\newblock in {\em International Conference on Unconventional Computation and
  Natural Computation}, June 2019, pp. 232--256.

\bibitem{annealer2019isscc}
T.~{Takemoto}, M.~{Hayashi}, C.~{Yoshimura}, and M.~{Yamaoka},
\newblock ``2.6 a 2 ×30k-spin multichip scalable annealing processor based on
  a processing-in-memory approach for solving large-scale combinatorial
  optimization problems,''
\newblock in {\em IEEE International Solid- State Circuits Conference},
  February 2019.

\bibitem{chou2019analog}
J.~Chou, S.~Bramhavar, S.~Ghosh, and W.~Herzog,
\newblock ``Analog coupled oscillator based weighted ising machine,''
\newblock {\em Scientific Reports}, , no. 14786, pp. 1--10, October 2019.

\bibitem{darpa2019ai_campain}
Defense Advanced Research~Projects Agency,
\newblock ``Ai next campaign,'' 2019. [Online]. Available:
  https://www.darpa.mil/work-with-us/ai-next-campaign.

\bibitem{analog_meet_digital}
M.~{Verhelst} and A.~{Bahai},
\newblock ``Where analog meets digital: Analog-to-information conversion and
  beyond,''
\newblock {\em IEEE Solid-State Circuits Magazine}, vol. 7, no. 3, pp. 67--80,
  September 2015.

\bibitem{glass1991daido}
H.~Daido,
\newblock ``Quasientrainment and slow relaxation in a population of oscillators
  with random and frustrated interactions,''
\newblock {\em Physical review letters}, vol. 68, no. 7, pp. 1073, February
  1992.

\bibitem{adler1947}
R.~{Adler},
\newblock ``A study of locking phenomena in oscillators,''
\newblock {\em Proceedings of the IRE}, vol. 34, no. 6, pp. 351--357, June
  1946.

\bibitem{vosoughi201610}
M.A. Vosoughi,
\newblock {\em A 10 BIT INTERFACE CIRCUIT FOR AN ARRAY OF CAPACITIVE
  TRANSDUCERS},
\newblock Ph.D. thesis, Bogazi{\c{c}}i University, 2016.

\bibitem{vosoughi2017noise}
M.A. Vosoughi, H.~Torun, and G.~Dundar,
\newblock ``Noise analysis in switched capacitor amplifier based sensors,''
\newblock in {\em New Generation of Circuits and Systems}. IEEE, 2017, pp.
  249--252.

\bibitem{vosoughi2019bus}
M~Ali Vosoughi, Longfei Wang, and Sel{\c{c}}uk K{\"o}se,
\newblock ``Bus-invert coding as a low-power countermeasure against correlation
  power analysis attack,''
\newblock in {\em 2019 ACM/IEEE International Workshop on System Level
  Interconnect Prediction (SLIP)}. IEEE, 2019, pp. 1--5.

\bibitem{vosoughi2019leveraging}
M.A. Vosoughi and S.~K{\"o}se,
\newblock ``Leveraging on-chip voltage regulators against fault injection
  attacks,''
\newblock in {\em Proceedings of the 2019 on Great Lakes Symposium on VLSI},
  2019, pp. 15--20.

\bibitem{ptm}
NIMO Group Arizona~State University,
\newblock ``Predictive technology model (ptm),'' 2008. [Online]. Available:
  http://ptm.asu.edu/.

\bibitem{frasca2018synchronization}
M.~Frasca \textit{et al.},
\newblock {\em Synchronization in Networks of Nonlinear Circuits: Essential
  Topics with MATLAB{\textregistered} Code},
\newblock Springer, March 2018.

\end{thebibliography}
